\documentclass[12pt, preprint]{aastex}
\begin{document}
\title{Discovery of Fast X-ray Oscillations During the 1998 Giant
Flare from SGR 1900+14} \author{Tod E. Strohmayer$^1$ and Anna
L. Watts$^2$} \altaffiltext{1}{Exploration of the Universe Div.,
NASA/GSFC, Greenbelt, MD 20771; stroh@milkyway.gsfc.nasa.gov}
\altaffiltext{2}{Exploration of the Universe Div., NRC/GSFC,
Greenbelt, MD 20771; anna@milkyway.gsfc.nasa.gov}
\begin{abstract}

We report the discovery of complex high frequency variability during
the August 27, 1998 giant flare from SGR 1900+14 using the Rossi X-ray
Timing Explorer (RXTE). We detect an $\approx 84$ Hz oscillation (QPO)
during a 1 s interval beginning approximately 1 min after the initial
hard spike.  The modulation amplitude is energy dependent, reaching a
maximum of 26\% (rms) for photons above 30 keV, and is not detected
below 11 keV, with a 90\% confidence upper limit of 14\% (rms).
Remarkably, additional QPOs are detected in the average power spectrum
of data segments centered on the rotational phase at which the 84 Hz
signal was detected. Two signals, at 53.5 and 155.1 Hz, are strongly
detected, while a third feature at 28 Hz is found with lower
significance. These QPOs are not detected at other rotational
phases. The phenomenology seen in the SGR 1900+14 flare is similar to
that of QPOs recently reported by \citet{isr05} from the December 27,
2004 flare from SGR 1806-20, suggesting they may have a common origin,
perhaps torsional vibrations of the neutron star crust.  Indeed, an
association of the four frequencies (in increasing order) found in SGR
1900+14 with $l$ = 2, 4, 7, and 13 toroidal modes appears
plausible. We discuss our findings in the context of this model and
show that if the stars have similar masses then the magnetic field in
SGR 1806-20 must be about twice as large as in SGR 1900+14, broadly
consistent with magnetic field estimates from pulse timing.

\end{abstract}

\section{Introduction}

The giant flares produced by highly magnetized neutron stars,
``magnetars,'' are amongst the most energetic of stellar
phenomena. Three such flares have been recorded since the advent of
satellite borne high energy detectors.  The most recent event, from
SGR 1806-20 (hereafter, 1806), was also the most luminous, generating
upwards of $10^{46}$ erg in a matter of minutes \citep{ter05,
pal05}. At its peak the prodigious photon flux paralyzed almost all
satellite instruments which viewed the event.  The magnetic
instability which powers such flares is likely associated with large
scale fracturing of the neutron star crust \citep{flo77, tho95, dun98,
td01, sch05}, and will almost certainly excite large scale seismic
motions within the star.

Recently, \citet{isr05} presented evidence from Rossi X-ray Timing
Explorer (RXTE) observations of transient, high frequency
quasi-periodic oscillations (QPOs) during the pulsating tail of the
giant flare from 1806.  They found 92.5 Hz QPOs approximately 200 s
after the onset of the flare.  These QPOs appear to be associated with
a spectrally hard, unpulsed emission component, and are only detected
during particular rotational phases. The presence of lower frequency,
$\approx 18$ and $30$ Hz QPO features was also claimed.  \citet{isr05}
suggested that these QPOs may be due to global oscillation modes of
the neutron star.  In particular, they propose an association with
torsional (toroidal) vibrations of the neutron star crust
\citep{sch83, mcd88, str91, dun98}.  Toroidal oscillations can be
readily excited by crust fracturing, as demonstrated by the
devastating 2004 Sumatra-Andaman earthquake \citep{par05}, and have
periods within the range observed. Coupling to the magnetosphere via
the magnetic field threading the crust appears to be a plausible
method of modulating the X-ray flux during the flare, but a detailed
model of this process is needed. We note that Barat et al. (1983)
suggested torsional vibrations might explain a claimed 23 ms
modulation during the initial spike of the March 5, 1979 flare from
SGR 0526-66.

Motivated by the work of \citet{isr05} we investigated the fast timing
properties of the only other giant flare observed with high
sensitivity instrumentation; the August 27, 1998 event from SGR
1900+14 (hereafter, 1900, Hurley et al. 1999). \citet{fer01} carried
out a detailed study of the {\it BeppoSAX} and {\it Ulysses} data from
this flare.  They found a clear signal of the neutron star's spin
frequency ($ \sim 0.2$ Hz) and its harmonics in power spectra of the
event, and modelled the pulsating tail with a trapped fireball model.
\citet{guid04} investigated the flare's fast timing properties with
{\it BeppoSAX}, but could not search the entire event due to a dearth
of high time resolution data. Nevertheless, they presented evidence
for a power law noise component extending above 100 Hz.

In this Letter we report the results of a timing study of this flare
using the Proportional Counter Array (PCA) onboard RXTE.  We detect a
transient QPO at $\approx 84$ Hz during a $\sim 1$ s interval
approximately 1 minute after flare onset, and a pair of persistent
QPOs at 53.5 and 155.5 Hz associated with the rotational phase of the
84 Hz signal.  After describing our timing study we conclude with a
discussion of our findings in the context of neutron star toroidal
oscillation modes.

\section{Observations and Data Analysis}

During the August 27, 1998 flare from 1900, RXTE was pointed at the
cataclysmic variable AM Her, about $42^{\circ}$ away from it.  The PCA
was configured in one of the ``good xenon'' data modes (Goodxenon16s)
with $\approx 1$ $\mu$s resolution.  This mode was identical to one
running at the time of the 1806 flare except for a longer read-out
cycle (16 s compared with 2 s for 1806) that led to data gaps during
the flare.  Nevertheless, high time resolution, unsaturated data are
available for portions of the pulsating tail.  Fig. 1 shows the
lightcurve of the flare from both the Standard1 (bottom) and
Goodxenon16s (top) modes. The Standard1 mode produces 1/8 s resolution
light curves across the full PCA bandpass (nominally 2 - 90 keV). The
gaps in the Goodxenon16s data occur because the buffer is filled
before it can be read out. The buffer can store 16,000 events, and is
read out every 16 s, so above a mean rate of $\approx 1000$ s$^{-1}$
gaps are inevitable.

Because of the gaps we first searched each good data interval
separately, starting with the first interval following the peak of the
flare (severe deadtime at the peak precludes a meaningful search
there).  We calculated power spectra with a Nyquist frequency of 2048
Hz using the good data in each interval.  Based on the results of
\citet{isr05} we averaged the power spectra to $\approx 1$ Hz
frequency resolution to best match their reported QPO coherence
values.  We did not find any candidate detections in the first three
data intervals, however, the power spectrum of the fourth interval
(marked by the vertical dotted lines in Fig. 1) shows a significant
peak at $\approx 84$ Hz. Fig. 2 shows the power spectrum from this
interval in the 10 -- 2048 Hz band at 1 Hz resolution. We estimated
the significance using a $\chi^2$ distribution with 16 degrees of
freedom, which is the distribution expected based on the number of
independent frequency bins averaged.  We fitted the observed
distribution of noise powers with a $\chi^2$ function and confirmed
that it is consistent with the expected noise distribution (see the
inset to Fig. 2).  The peak at 84 Hz has a single trial probability of
$6.8 \times 10^{-10}$ based on this noise distribution.  Multiplying
by the number of frequency bins searched (2048) and a ``confirmation
bias'' reflecting the fact that this was the fourth spectrum searched
we arrive at a significance of $5.6 \times 10^{-6}$, a fairly robust
detection.  The peak power corresponds to an average amplitude (rms)
of 10.2\%.

We next constructed a dynamic power spectrum using the $Z^2$ statistic
\citep{str99}. Fig. 3 shows a contour plot of this spectrum and the
light curve of the fourth data segment.  The 84 Hz signal is clearly
localized in time near the largest peak of the four-peaked profile
that emerged during the pulsating tail \citep{hur99, fer01}. Although
the signal is {\it not} centered on the peak, we cannot rule out some
association with its right flank. This behavior is similar to that
reported by \citet{isr05}, in that the oscillations are not centered
on a peak.  To quantify the amplitude variation with energy we
computed power spectra for different energy bands, restricting the
time range to the interval in which the signal was detected. The
amplitude measurements are corrected for background, using data prior
to the flare as an estimate of the total background. The amplitude
(rms) in the 12 -- 90 keV band is $20\pm 3\%$, increasing to $26\pm
4\%$ for photons above 30 keV.  For photon energies less than 18 keV
we have a 90\% confidence upper limit of about 14\%. We note that a
detailed spectral study of the flare is complicated by the fact that
1900 was not in the PCA field of view.


Israel et al. (2005) found that the QPOs in 1806 had a dependence on
rotational phase.  To test for such behavior we computed an average
power spectrum using 1.5 s intervals centered on and beginning at the
rotational phase of the 84 Hz signal. For the spin period we used the
results of Woods et al. (1999). The resulting power spectrum from 20
independent intervals is plotted in Fig. 4 (middle and upper curves),
and clearly shows an additional pair of QPOs, and perhaps a third.
The time interval of this spectrum extends about 90 s beyond the 84 Hz
detection. We fit this spectrum with a power law continuum and a pair
of Lorenztian profiles for the two strongest QPOs.  This model fits
well, with a $\chi^2 / dof = 904/892$. The frequencies of the
strongest QPOs are $53.5 \pm 0.5$ and $155.1 \pm 0.2$ Hz, with
coherence values $Q \equiv \nu/\sigma_{\nu}$ of 22 and 55,
respectively. Their average amplitudes (rms) are $6.7 \pm 1.5$ and
$10.9 \pm 2.2$ \%. These QPOs are very significant; removal of the 53
and 155 Hz components increases $\chi^2$ by 72 and 134, respectively.
The middle and top curves in Fig. 4 show the same power spectrum at
different frequency resolution, 4 and 4/3 Hz, respectively. At the
higher frequency resolution (top) there is a suggestive peak at about
28 Hz. Including a Lorentzian for this feature improves $\chi^2$ by
22.1, which is a bit better than a $3\sigma$ deviation, so we consider
this a tentative detection.  Finally, the strength of these features
is a function of rotational phase. To confirm this we computed average
power spectra in the same manner but centered on rotational phases
{\it away} from that of the 84 Hz signal. Fig. 4 (bottom) shows the
average power spectrum for phases centered $\pm 1.5$ s from that of
the 84 Hz signal. The QPOs are clearly not detected. This analysis
also confirms that the detected signals cannot be detector artifacts,
but must be intrinsic to the source.

\section{Discussion}

Our results show striking similarities with those obtained for the
1806 flare by \citet{isr05}.  First, we find a frequency within 8\% of
the 92.5 Hz feature in 1806, suggesting the same underlying
mechanism. Second, the strongest signal is transient, appears to be
associated with a particular rotational phase and is not centered on a
profile peak. Although the 92 Hz signal in 1806 persists for longer,
this could be due to its larger overall luminosity, and we may have
missed occurrences of the 84 Hz signal due to data gaps.  In both
flares the amplitude is greater at higher photon energies. However,
there is no obvious hard DC component associated with the 84 Hz QPO in
1900, as in 1806 \citep{isr05}.

The firm detection of multiple QPOs associated with a specific
rotational phase provides compelling evidence that the modulations are
related to a particular region on the stellar surface, perhaps the
site of a crust fracture, or a certain magnetic field bundle.
Moreover, the comparisons to 1806 strongly suggest we are seeing a
similar phenomenon in 1900.  \citet{isr05} conjectured that toroidal
oscillations of the neutron star might explain the observed
frequencies. We now explore the implications of our findings in the
context of this model.

For a non-rotating, non-magnetic star, \citet{dun98} estimates the
period of the fundamental ($l=2, \; n=0$) toroidal mode (denoted $_2
t_0$) to be
\begin{equation}
P(_2 t_0) = 33.6 R_{10} \frac{0.87 + 0.13 M_{1.4}
      R_{10}^{-2}}{(1.71 -
0.71 M_{1.4} R_{10}^{-1})^{1/2}}~~\mathrm{ms}  \; ,
\label{e1}
\end{equation}
where $R_{10} = R/10$km and $M_{1.4}=M/1.4M_\odot$. 
The periods of higher order $n=0$ modes ($n \ge 1 $
modes have frequencies that are too high) are given by
\begin{equation}
P(_l t_0) = P(_2 t_0) \left( \frac{6}{l(l+1)}\right)^{1/2} \left[1 +
\left(\frac{B}{B_\mu}\right)^2\right]^{-1/2} \; ,
\label{e2}
\end{equation}
where the factor in square brackets is a magnetic correction, $B_\mu
\approx 4 \times 10^{15} \rho_{14}^{0.4}$G, and $\rho_{14}$ is the
crustal density in units of $10^{14}$ g cm$^{-3}$. \citet{dun98}
argues that $\rho_{14} \sim 1$ is appropriate since the mode periods
are largely set by conditions in the deep crust. In deriving equation
(\ref{e2}) it is assumed that magnetic tension boosts the crustal
shear modulus isotropically.  However, the field configuration and
corresponding non-isotropic effects could alter the magnetic
correction dramatically \citep{mes01}.

\citet{isr05} suggest that their 92.5 Hz QPO, and a weaker feature at
30.4 Hz could be due to the $_7t_0$ and $_2t_0$ modes, respectively.
Preferential excitement of higher $l$ modes is not unreasonable, being
set by the fracture geometry: such an effect was seen in the spectrum
of Earth modes excited by the 2004 Sumatra-Andaman earthquake
\citep{par05}. With at least three frequencies in 1900 we can ask if
identification with a sequence of modes of different $l$ is plausible.
We find only one reasonably plausible sequence, that where the
frequencies are due to the $l=4$ (53 Hz), $l=7$ (84 Hz) and $l=13$
(155 Hz) modes (see Figure 4). With this sequence the inferred $_2t_0$
mode frequency is $\approx 28$ Hz, lower than that for 1806, but,
intriguingly, consistent with the frequency of our tentatively
detected 28 Hz QPO. Although the mode frequencies in this sequence do
not precisely match our QPO centroid frequencies, magnetic field
corrections could plausibly account for the modest offsets.

The different inferred $l = 2$ mode frequencies imply that the stellar
parameters must differ.  Given an equation of state (EOS), equations
(\ref{e1})-(\ref{e2}) specify the relationship between mass and
magnetic field necessary to give an oscillation at a desired
frequency.  Fig. 5 shows, for several EOS, the stellar parameters that
give $_2t_0$ oscillations at 28 Hz (1900) and 30.4 Hz (1806).  If the
stars have similar magnetic field strengths, their masses must differ
by more than $0.2 M_\odot$.  This seems unlikely unless magnetar
masses differ significantly from those of other non-accreting neutron
stars; a study of radio pulsars by \citet{tho99} found the masses to
be consistent with a narrow Gaussian distribution, $M=1.35\pm 0.04
M_\odot$.  More likely the stars have similar masses but different
magnetic fields.  If we assume that both stars have masses $\approx
1.35 M_\odot$, we can immediately rule out the softest EOS; and the
hardest EOS predicts magnetic fields for both systems that are far
larger than those inferred from timing studies \citep{woo02}.  Neither
the AP3 nor the AP4 EOS can be ruled out, given that the inferred
$l=2$ mode frequency of 28 Hz has a $\pm 0.5$ Hz uncertainty.
However, the magnetic fields that one would infer from these EOS agree
well with those derived from timing measurements of both stars, which
suggests that 1806 has a field strength approximately double that of
1900 \citep{woo05}.

\citet{fer01} and \citet{td01} have explored the radiative mechanisms
associated with magnetar flares. They associate the four-peaked
profile during the 1900 flare with X-ray ``jets'' produced by the
strong polarization dependence of scattering in the $\approx 10^{15}$
G field. These jets form close to the surface and likely reflect the
multipole structure of the magnetic field. Our detection of QPO
signals shortly after the emergence of the four-peaked profile, in
association with one of these ``jets'', is consistent with a mechanism
that produces motions at or near the stellar surface. Indeed, the
presence of beamed emission offers a plausible scenario to achieve
large amplitudes from modest surface displacements. For example,
perturbations to the magnetic field geometry could produce changes in
the jet angle with respect to a distant observer, creating a large
change in observed flux.



\section{Summary}

The discovery of high frequency QPOs with the right $[l(l+1)]^{1/2}$
scaling during a second magnetar flare is strong evidence that we may
be seeing the influence of global toroidal modes of neutron star
crusts on the X-ray emission from these objects.  The recent results
suggest these QPOs may be ubiquitous, in which case future
observations of such flares, which we know will occur, albeit
infrequently, could provide a new sensitive probe of neutron star
structure.  Efforts should be made to see that sensitive
instrumentation is in place to observe the next such flare.


\acknowledgements

We thank Padi Boyd, Craig Markwardt, Jean Swank, Michael Rupen,
Nikolai Shaposhnikov, Stephen Holland, Sudip Bhattacharyya, and the
referee for comments and discussions that helped stimulate this work.

\clearpage

\begin{figure}
\begin{center}
\includegraphics[width=6in, height=6in, clip]{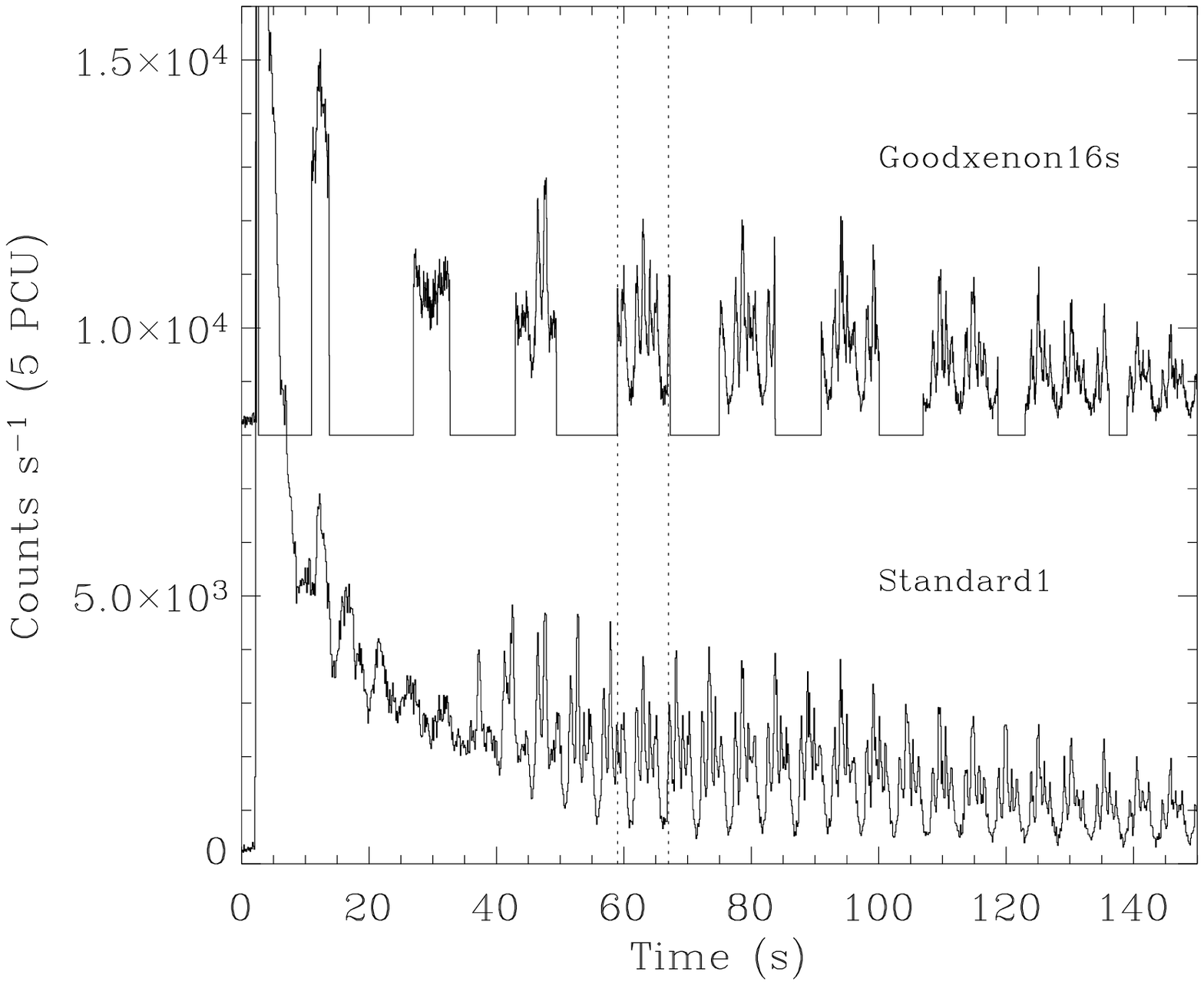}
\end{center}
\caption{Light curves of the August 27, 1998 flare from 1900.  The
Goodxenon16s curve (top) is plotted at 1/16 s resolution, and is
displaced vertically by 8,000 s$^{-1}$ for clarity. The data gaps
result from the finite buffer size and the long time between read outs
(16 s).  The Standard1 data (bottom) provides an uninterrupted view of
the flare. The vertical dotted lines mark the time interval in which
the 84 Hz signal was first detected. Time zero corresponds to MJD
51052.33088176 (TT).}
\end{figure}

\pagebreak

\begin{figure}
\begin{center}
\includegraphics[width=6in, height=6in, clip]{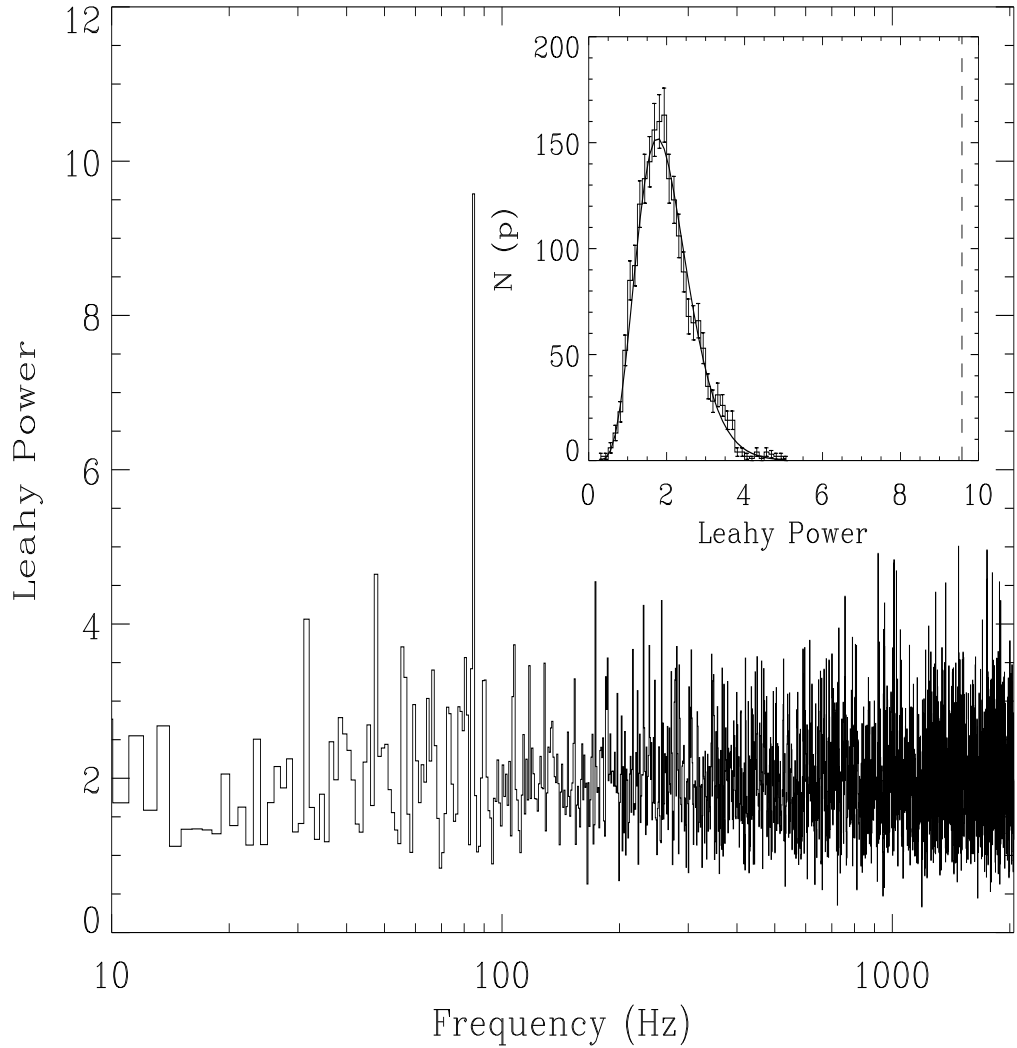}
\end{center}
\caption{Power spectrum in the 10 -- 2048 Hz band of the fourth good
data interval from the Goodxenon16s data (see Fig. 1), showing the 84
Hz peak. The frequency resolution is 1 Hz, and the Nyquist frequency
is 2048 Hz. The inset panel shows the distribution of powers as a
histogram, along with the best fitting $\chi^2$ function (solid). The
peak power in the 84 Hz feature is marked with the vertical dashed
line.}
\end{figure}

\pagebreak

\begin{figure}
\begin{center}
\includegraphics[width=6in, height=6in, clip]{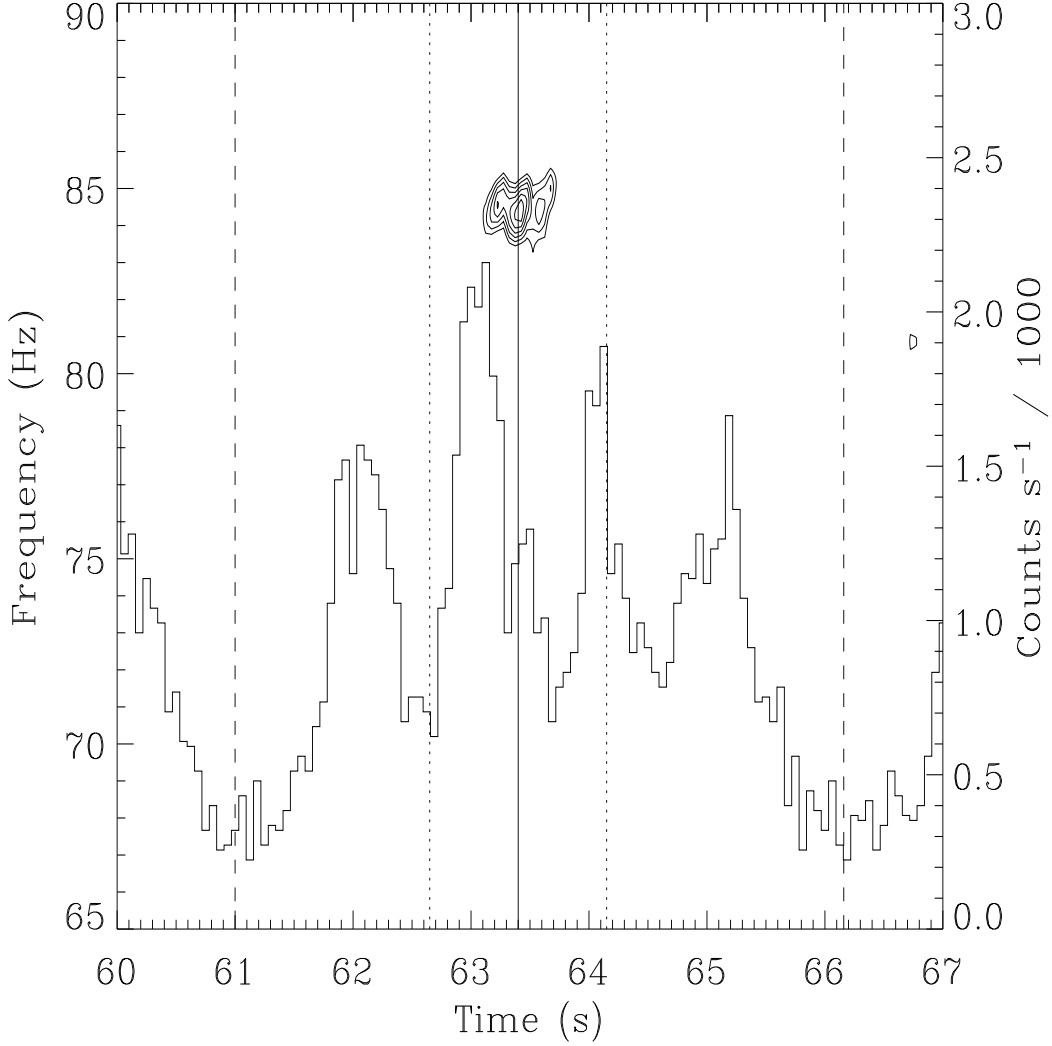}
\end{center}
\caption{Dynamic power spectrum of the data interval containing the
QPO. Spectral power contours were computed from 0.5 s segments with a
new segment beginning every 0.1 s. The lowest and highest contours
correspond to single trial probabilities of $4.5 \times 10^{-5}$ and
$1.7\times 10^{-10}$, respectively. The signal is localized near the
brightest peak of the four-peaked profile, but it is not centered on
the peak. The vertical dotted lines denote the phase range used to
compute average power spectra (see Fig. 4). The vertical solid line
marks the center of this range. One rotational cycle (5.16 s period)
is marked by the vertical dashed lines. The time scale is the same as
in Fig. 1.}
\end{figure}

\pagebreak

\begin{figure}
\begin{center}
\includegraphics[width=6in, height=6in,clip]{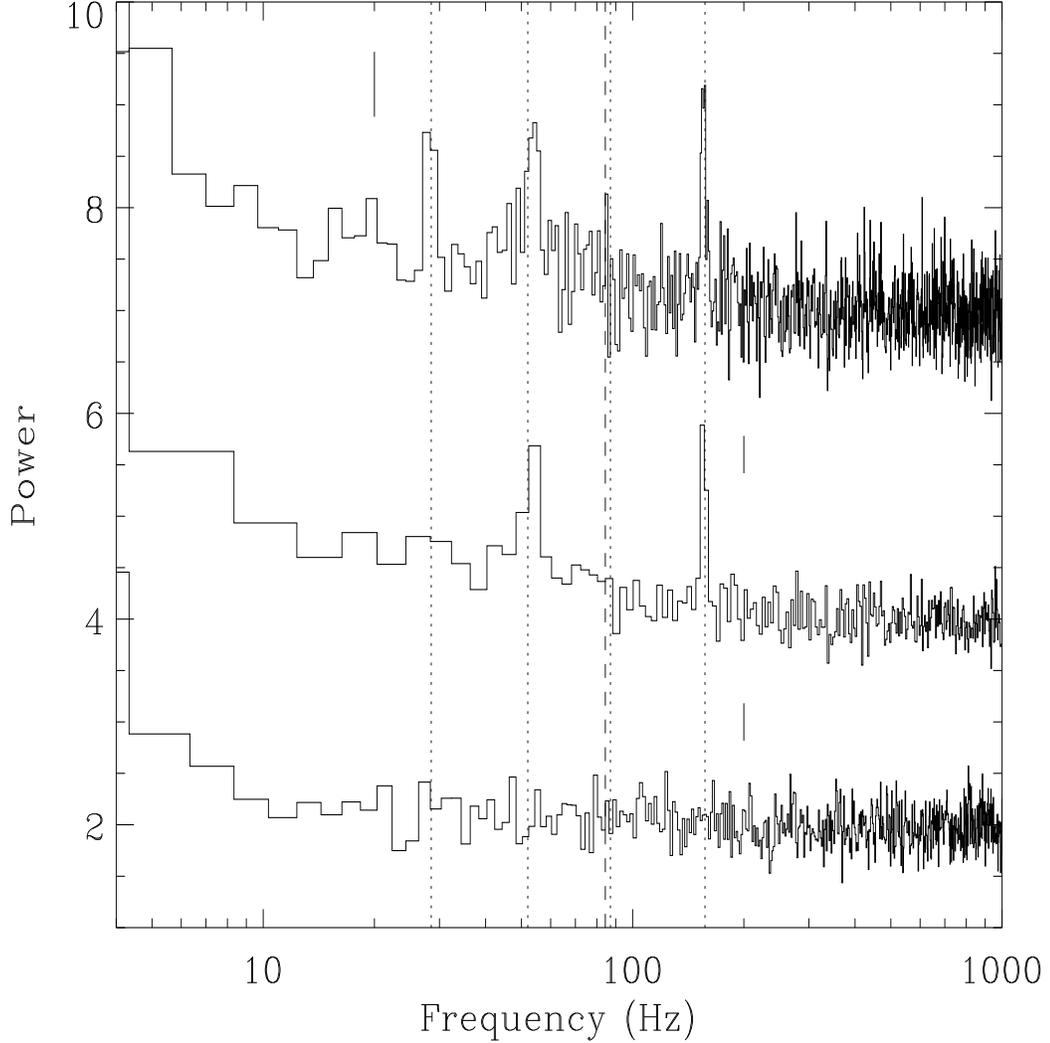}
\end{center}
\caption{Average power spectra from 1.5 s intervals centered at
different rotational phases.  The upper curve was computed using 20
successive 1.5 s intervals centered on the rotational phase of the 84
Hz QPO (see Fig. 3). The frequency resolution is 4/3 Hz.  The middle
curve shows the same spectrum with 4 Hz resolution.  Two QPOs are
clearly detected in this spectrum, and a third, weaker feature is
present at the higher frequency resolution (top, see the text for
details). The lower curve is from reference phases $\pm 1.5$ s away
from the 84 Hz signal phase. No QPOs are detected at these phases. The
vertical dotted lines mark the frequencies of $l$ = 2, 4, 7, and 13
$_lt_0$ modes, in order of increasing frequency, assuming that the
$_2t_0$ mode has a frequency of 28.5 Hz. The vertical dashed line
marks the frequency of the transient 84 Hz QPO. At the 5.16 s period,
1.5 s corresponds to 0.3 cycles, or about $105^{\circ}$. Characteristic 
error bars are shown for each spectrum.}
\end{figure}

\pagebreak

\begin{figure}
\begin{center}
\includegraphics[width=6in, height=6in, clip]{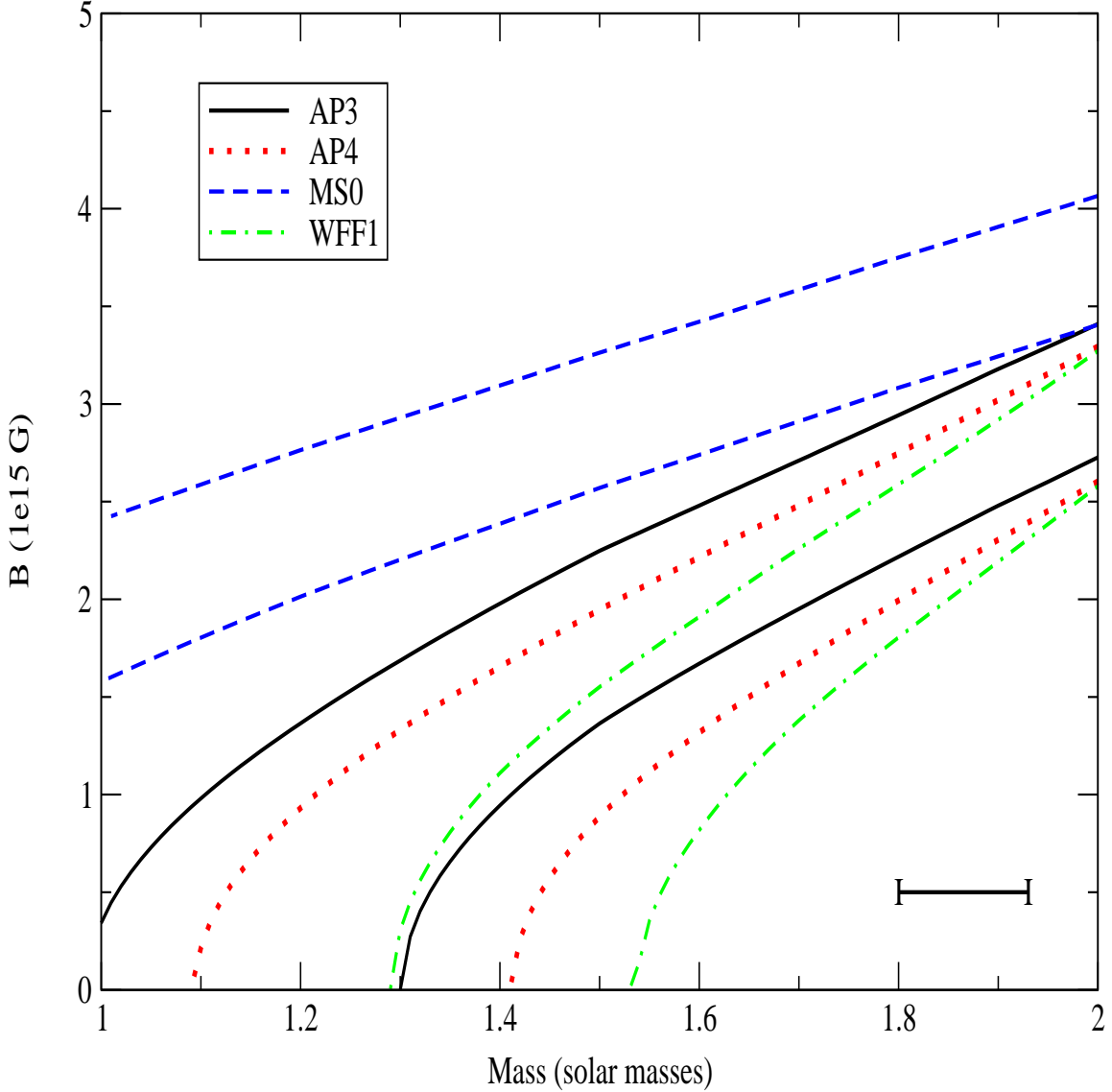}
\end{center}
\caption{The stellar mass and magnetic field required to give the
 $_2t_0$ mode frequencies inferred for 1806 (30.4 Hz) and 1900
 ($\approx$ 28 Hz).  We model four of the EOS discussed in
 \citet{lat01}; in order of increasing stiffness they are WFF1, AP4,
 AP3 and MS0.  For each EOS two lines are shown.  The upper line
 indicates the parameters necessary to give 30.4 Hz, the lower line 28
 Hz.  The horizontal line indicates the uncertainty in the position of
 the footprint of the 1900 AP3, AP4 and WFF1 lines due to the $\pm $
 0.5 Hz uncertainty on the 28 Hz frequency.}
\end{figure}


\begin{thebibliography}{}

\bibitem[Barat(1983)]{Bar83} Barat, C. et al. 1983, A\&A, 126, 400
\bibitem[Duncan(1998)]{dun98} Duncan, R.C., 1998, ApJ, 498, L45
\bibitem[Feroci et al.(2001)]{fer01} Feroci, M., et al., 2001, ApJ, 549, 1021
\bibitem[Guidorzi et al.(2004)]{guid04} Guidorzi, C. et al., 2004, A\&A,
416, 297
\bibitem[Flowers \& Ruderman(1977)]{flo77} Flowers, E., Ruderman,
M.A., 1977, ApJ, 215, 302
\bibitem[Hurley et al.(1999)]{hur99} Hurley, K., et al., 1999, Nature,
  397, 41
\bibitem[Israel et al.(2005)]{isr05} Israel, G., et al., 2005, \apj,
  628, L53
\bibitem[Lattimer \& Prakash(2001)]{lat01} Lattimer, J.M., Prakash,
  M., 2001, ApJ, 550, 426
\bibitem[McDermott, van Horn \& Hansen(1988)]{mcd88} McDermott, P.N.,
  van Horn, H.M., Hansen, C.J., 1988, ApJ, 325, 725
\bibitem[Messios, Papadopoulos \& Stergioulas(2001)]{mes01} 
Messios, N., Papadopoulos, D.B., Stergioulas, N., 2001, MNRAS, 328, 1161
\bibitem[Palmer et al.(2005)]{pal05} Palmer, D.M., et al., 2005,
Nature, 434, 1107
\bibitem[Park et al.(2005)]{par05} Park, J., et al., 2005, Science, 308, 1139
\bibitem[Schumaker \& Thorne(1983)]{sch83} Schumaker, B.L., Thorne,
  K.S., 1983, MNRAS, 203, 457
\bibitem[Schwartz et al.(2005)]{sch05} Schwartz, S.J., et al., 2005,
  \apj, 627, L129
\bibitem[Strohmayer(1991)]{str91} Strohmayer, T.E., 1991, ApJ, 372, 573
\bibitem[Strohmayer \& Markwardt(1999)]{str99} Strohmayer, T.E.,
  Markwardt, C.B., 1999, ApJ, 516, L81
\bibitem[Terasawa et al.(2005)]{ter05} Terasawa, T., et al., 2005,
  Nature, 434, 1110 
\bibitem[Thompson \& Duncan(2001)]{td01} Thompson, C., \& Duncan,
R. C. 2001, ApJ, 561, 980
\bibitem[Thompson \& Duncan(1995)]{tho95} Thompson, C., Duncan, R.C.,
  1995, MNRAS, 275, 255 
\bibitem[Thorsett \& Chakrabarty(1999)]{tho99} Thorsett, S.E.,
  Chakrabarty, D., 1999, ApJ, 512, 288 
\bibitem[Woods et al.(2002)]{woo02} Woods, P.M., et al., 2002, ApJ, 576, 381
\bibitem[Woods et al.(1999)]{woo99} Woods, P.M. et al. 1999, ApJ, 524, L55
\bibitem[Woods \& Thompson(2005)]{woo05} Woods, P.M., Thompson, C., to
  appear in {\it Compact Stellar X-ray Sources}, eds. W.H.G. Lewin and
  M. van der Klis, Cambridge University Press
\end{thebibliography}
\end{document}